\documentclass{aa}  
\usepackage{graphicx}

\usepackage{txfonts}
\usepackage{hyperref}
\def\h2{\hskip-2pt}

\begin{document}

   \titlerunning{Separation ratios}

   \title{On the use of the ratio of small to large separations in  asteroseismic model fitting }

   \author{Ian W. Roxburgh and Sergei V Vorontsov}

   \institute{Astronomy Unit, Queen Mary University of London, 
     Mile End Road, London E1 4NS, UK.
   \email {I.W.Roxburgh@qmul.ac.uk} }

   \date{Received  / Accepted }

 
    \abstract
 {The use of ratios of small to large separations as a diagnostic of stellar interiors.} 
{To demonstrate that model fitting by comparing observed and model separation ratios at the same  $n$  values  is in error, and to present a correct procedure.}
{Theoretical analysis using phase shifts and numerical models .}
{We show that the separation ratios of stellar models with the same interior structure, but different outer layers, are not the same when compared at the same $n$ values, but are the same when evaluated at the same frequencies by interpolation.  The separation ratios trace the phase shift differences as a function of frequency not of $n$. We give examples from model fitting where the ratios at the same $n$ values agree within the error estimates, 
but do not agree when evaluated at the same frequencies and the models do not have the same interior structure.  The correct procedure is to compare observed ratios with those of models interpolated to the observed frequencies.}
  {}

   \maketitle
%
\section{Introduction}
The ratio of small to large separations of stellar p-modes as a diagnostic  of the internal structure of stars was proposed in  Roxburgh \& Vorontsov (2003)  (see also Roxburgh 2004, 2005, Ot{\'i}  Floranes et al 2005) and is increasingly being used in model fitting, that is finding that model (or models), out of a set models, whose separation ratios best fit the the ratios of the observed frequencies.  These ratios, defined in terms of the frequencies $\nu_{n\ell}$  as (Roxburgh \& Vorontsov 2003) 
$$r_{02}(n) = {\nu_{n.0} - \nu_{n-1,2}\over \nu_{n,1}-\nu_{n-1,1}}~~~~~r^*_{01}(n) = {\nu_{n,0} -( \nu_{n-1,1} +\nu_{n,1})/2\over \nu_{n,1}-\nu_{n-1,1}}$$
$$r_{01}(n)={(\nu_{n-1,0}-4\nu_{n-1,1}+6\nu_{n,0}-4\nu_{n,1}+\nu_{n+1,0})\over 8\, (\nu_{n,1}-\nu_{n-1,1})}\eqno(1)$$
(and  similarly defined ratios $r_{13}, r^*_{10}, r_{10}$)  subtract off the major contribution of the outer layers of a star and so constitute a diagnostic of the stellar interior, 
which is independent of the structure of the outer layers which are subject to considerable uncertainties in modelling, 
 
This validity of this procedure rests on the result that that the contribution of the outer layers of a star to the oscillation frequencies $\nu_{n\ell}$ is very nearly independent of the degree $\ell$; the separation ratios are combinations of frequencies that seek to subtract off this $\ell$ independent contribution.

We here point out that a seeking a best fit model by direct comparison of observational and model ratios at the same \{$n,\ell$\} values,  e.g. searching for  minimum in the reduced  $\chi^2_n$  where
$$\chi^2_{n} =  {1\over N} \sum_{\ell} \sum_{n}  \left( r^{obs}_{0\ell}(n) - r^{mod}_{0\ell}(n)\over \sigma^r_{0\ell}(n)\right)^2\eqno(2)$$
 is, in principle, incorrect, and can give misleading, indeed incorrect,  results. Here $\sigma^r_{n\ell}(n)$ are the error estimates on the ratios derived from $\sigma_{n\ell}$ the error estimates on the frequencies $\nu_{n\ell}$, and $N$  the number of ratios. We show below that the correct procedure is to compare the observed ratios at the observed frequencies with the model ratios interpolated to the observed frequencies.

\section{Examples}
To illustrate this we take a model  ``observed" star of mass $1.2M_\odot$, radius $1.34R_\odot$
luminosity $2.45L_\odot$,  effective temperature $T_{eff}=6242^o$K, and central hydrogen  abundance $X_c=0.30$.
We assume a total of 33 ``observed" frequencies ($11\,n$ values, $\ell=0,1,2$) in the range $1534-2550\mu$Hz and an error on the frequencies of  $\sigma_\nu=0.2\mu$Hz; the mean large separation
 $\Delta=<\nu_{n,0}-\nu_{n-1,0}>=96.2\mu$Hz.

We then constructed two comparison models: \\
1.  A "scaled model" with the same dimensionless interior structure $\rho/(M/R^3)$, pressure $P/(GM^2/R^4)$ and adiabatic exponent $\Gamma_1$ as functions of the dimensionless radius $x=r/R$ as the ``observed' star but with a mass of  $1.35M_\odot$, and radius of $1.25R_\odot$, this model has a mean large separation of $\sim 114\mu$Hz\\
2.  Model99 which has  the same interior structure as our ``observed" star  $\rho(r), P(r), \Gamma_1(r)$ but is truncated at a radius $R=0.99R_o$ where $R_o$ is the radius of the ``observed" star; this model has   a mean large separation $\Delta=112\mu$Hz.

The interior structure of the observed star and the 2 comparison models are shown in Fig. 1. Of course the structure of model99 is identical to that of the observed star out to $r=0.99R_o$.

\begin{figure}[t]
  \begin{center} 
   \includegraphics[width=8.87cm]{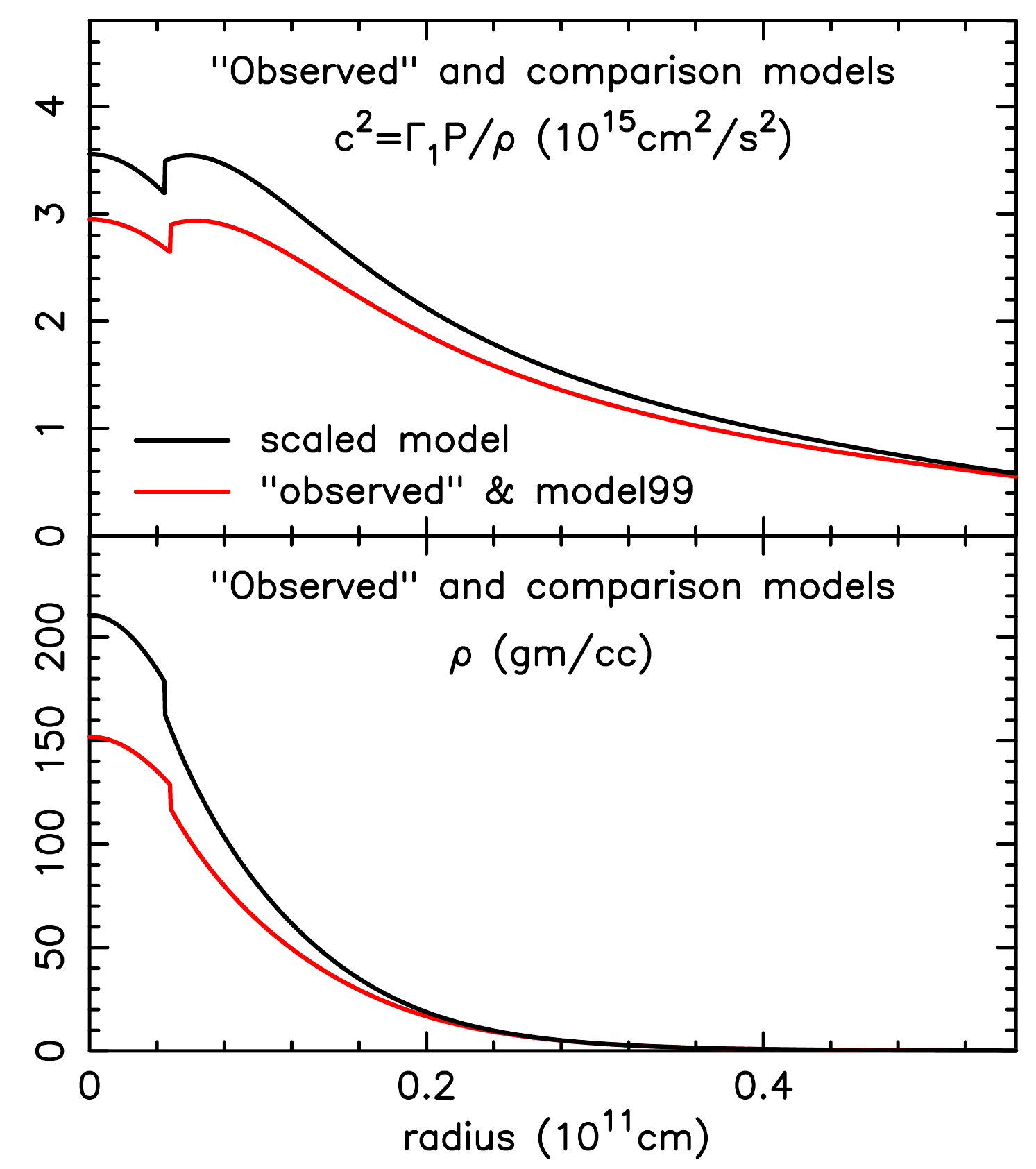}
  \end{center}  
   \vskip -15pt
   \caption{Internal structure  of the ÒobservedÓ and model stars}
    \begin{center} 
   \includegraphics[width=8.87cm]{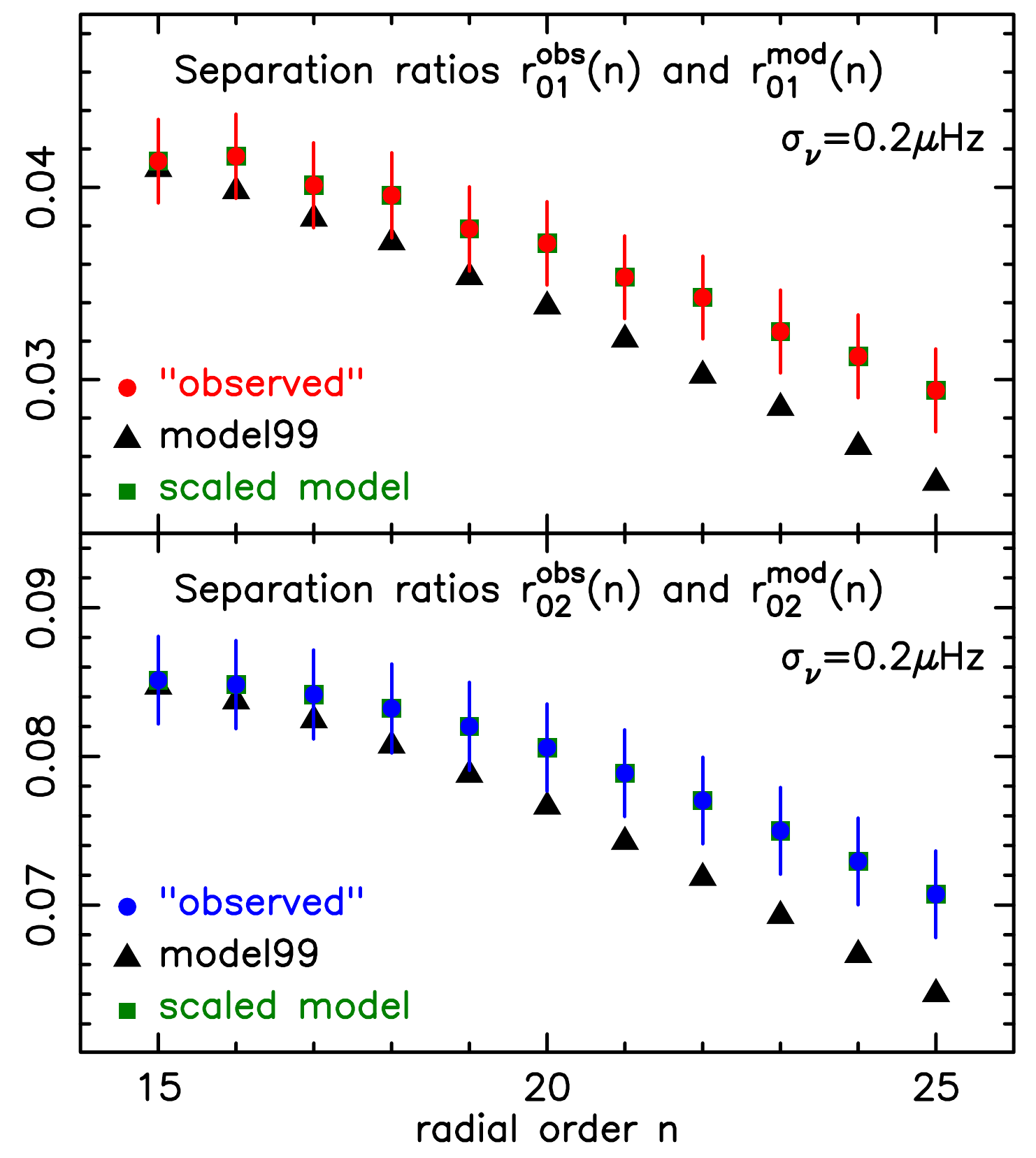}
  \end{center}  
   \vskip -15pt
   \caption{Top panel: the separation ratios $r_{01}(n)$ for model99 (triangles) and scaled model (squares) superimposed on those for the ``observed" star;
   bottom panel: the same but for the $r_{02}$ ratios  }
\end{figure}

In Fig. 2 we show the separation ratios $r_{0\ell}(n)$, as a function of radial order $n$,  the values for the observed and scaled model are exactly equal whereas those of model99 are substantially different even though its internal structure is identical to that of the observed model. Taking an error estimate on the frequencies of $\sigma_\nu=0.2\mu$Hz  gives a $\chi^2_n$ (as in Eqn. 2) of $2.2$ which  for 20 ratios has a probability of being due to chance of $0.15\%$. 

In Fig. 3 we plot the ratios $r_{0\ell}(n)$ against frequencies $\nu_{0n}$ for the ``observed" and comparison models. The scaled model does not agree with the observed model, whereas the truncated model values, whilst not evaluated at the same frequencies as the observed model, lie on a cubic interpolated  curve through the observed values.     
Comparing ratios at the same $n$ values (Eqn. 2) therefore tests the functional form of the dimensionless  structure of the stellar interior but not the physical values nor the mass and radius.  If the scaled model has the same $M/R^3$ as the observed star then the $r_{0\ell}(n)$ would coincide since the frequencies are invariant under a scaling that keeps $M/R^3$ constant.    

\begin{figure}[t]
  \begin{center} 
  \hskip 10pt
   \includegraphics[width=8.87cm,height=10cm]{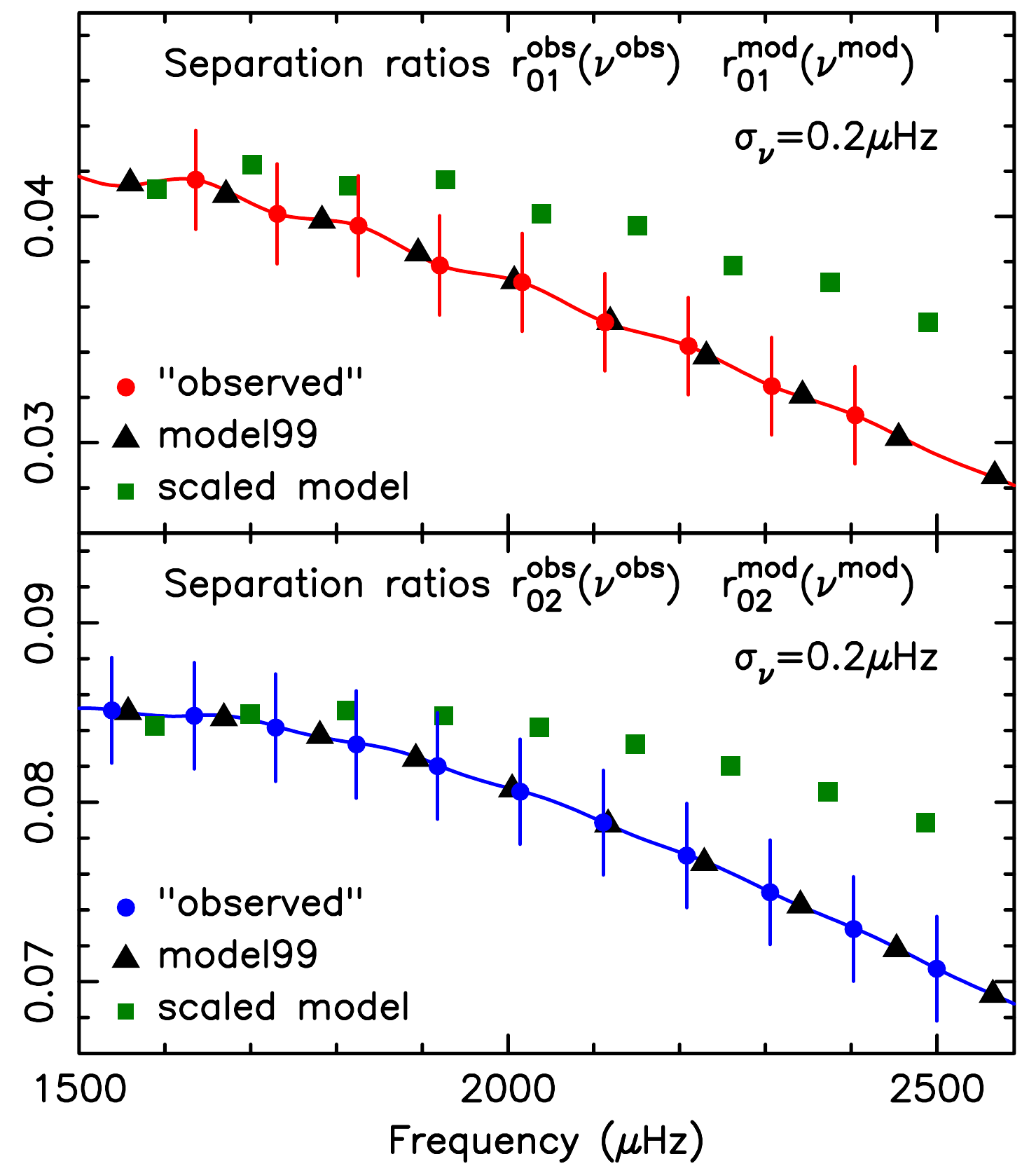}
  \end{center}  
   \vskip -15pt
   \caption{Top panel: the separation ratios $r_{01}(\nu)$ for model99 (triangles) and scaled model (squares) superimposed on those for the ``observed" star;
   bottom panel: the same but for the $r_{02}$ ratios  }
  \vskip -2pt
\end{figure}

\section{Separation ratios and phase shifts {\boldmath{$\epsilon(\nu)$}}}

The eigenfrequencies of a slowly rotating star can always be expressed  in the form
$$\nu_{n\ell} ={\bar  \Delta}\,\big[ n + \ell/2 + \epsilon_\ell(\nu_{n\ell}) \big] \eqno(3)$$ 
where ${\bar \Delta}$ is some estimate of the mean large separation, and the  {\it phase shifts} 
$\epsilon_\ell(\nu_{n\ell})$ are determined at the eigenfrequencies from this equation. 
The
$\epsilon_\ell(\nu_{n\ell})$ contain all the information on the departure of the structure of the star from a uniform sphere whose frequencies are the roots of the spherical  Bessel functions which
rapidly approach ${\bar \Delta}\, [n + \ell/2 ]$ as $n$ increases (Rayleigh 1894).

As shown in Roxburgh and Vorontsov (2000, 2003) (see section 6 below) the $\epsilon_\ell(\nu)$ can be effectively split into 
an $\ell$ dependent  inner contribution  determined by the structure of the interior of the star, and an (almost) $\ell$ independent  outer contribution determined by the structure of the outer layers.  Interpolating in
 $\epsilon_\ell(\nu)$,  which is known at the frequencies $\nu_{n\ell}$, to determine the value at any $\nu$, and subtracting  values for different $\ell$, cancels out both the $\ell$ independent contribution from the outer layers and the linear term $\nu/{\bar \Delta}$; the differences  $\epsilon_0(\nu) - \epsilon_\ell(\nu)$ as a  function of $\nu$  being determined solely by the structure of the inner layers  The separation ratios
 $r_{0\ell}$ 
 are approximations to values of $\epsilon_0(\nu) - \epsilon_\ell(\nu)$ at the frequencies $\nu_{n0}$.
 
\begin{figure}[t]
  \begin{center} 
   \includegraphics[width=8.87cm]{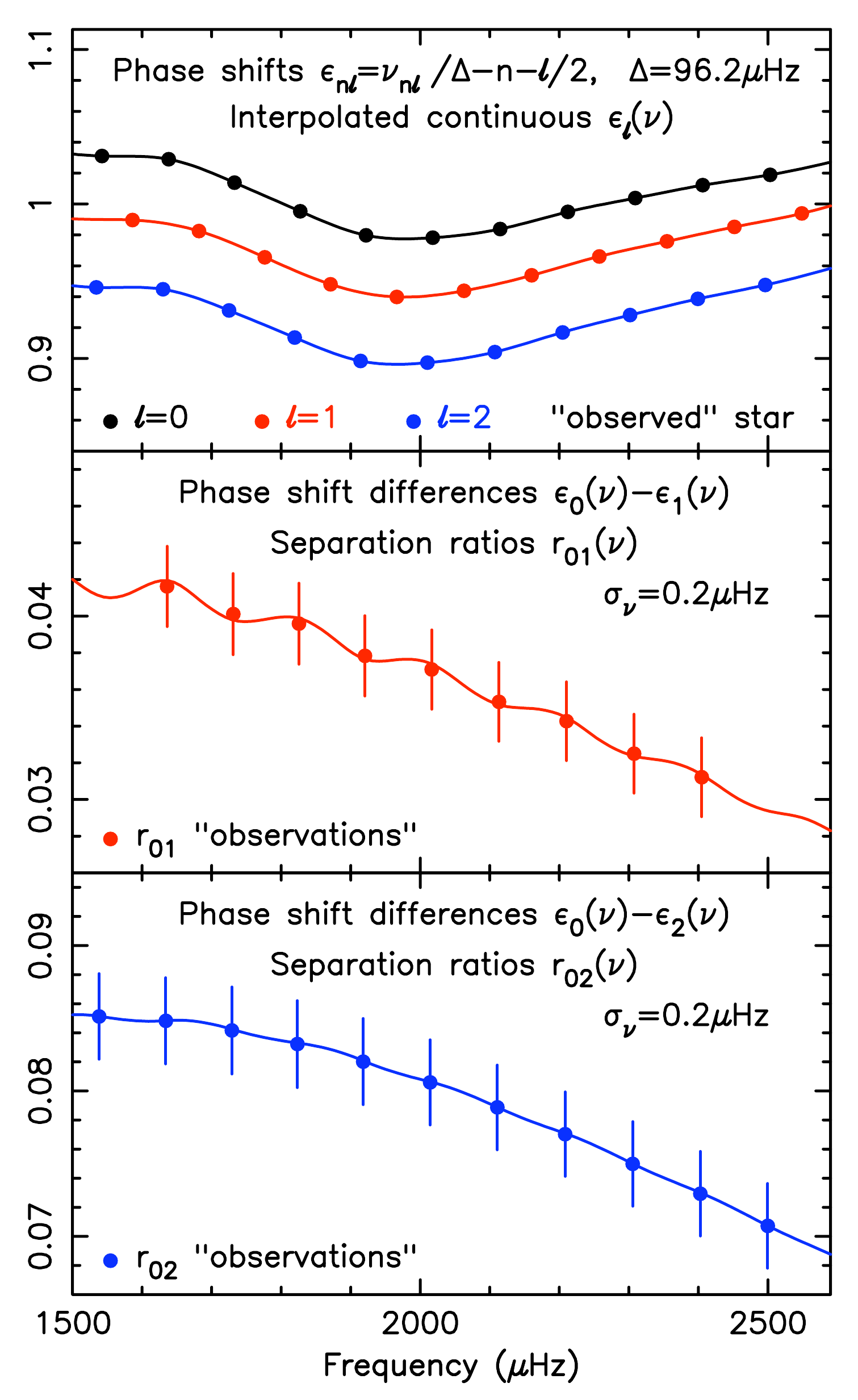}
  \end{center}  
   \vskip -12pt
   \caption{Top  panel:  phase shifts $\epsilon_\ell(\nu_{n\ell})$, and interpolated curves $\epsilon_\ell(\nu)$; middle panel: phase shift differences 
$\epsilon_0(\nu)-\epsilon_1(\nu)$ and the 9 separation ratios $r_{01}$ : bottom panel as above but for the 11 $r_{02}$.}
  \vskip-10pt
\end{figure}
 
The top panel in Fig. 4 shows the phase shifts  $\epsilon_\ell(\nu_{n\ell})$ at the observed frequencies for our ``observed" star, and the cubic interpolated curves 
$\epsilon_\ell(\nu)$, the bottom panels show the phase shift differences $\epsilon_0(\nu)-\epsilon_\ell(\nu)$ together with the separation ratios $r_{01}, r_{02}$ which lie on or close to the  curves $\epsilon_0(\nu)-\epsilon_\ell(\nu)$

If we now compare the separation ratios of star with observed frequencies $\nu_{n\ell}^{obs}$ to those of a model with different frequencies $\nu_{n\ell}^{mod}$ (as in Eqn.  2)  then we are comparing the values of the phase shift differences $\epsilon_\ell(\nu)-\epsilon_0(\nu)$ at different frequencies so, even if the star and model have the same interior structure, that is the same $\epsilon_\ell(\nu)-\epsilon_0(\nu)$ as a function of $\nu$, they will not have the same values of the separation ratios at their respective \{$n,\ell$\} values. This explains why the ratios for model99 in Fig. 3 are offset from the ``observed" values. The scaled model on the other hand has a different interior structure and therefore a different function for its phase shift differences.

In summary the separation ratios of model99 with the same interior structure as the ``observed" star lie on the same     $\epsilon_0(\nu) - \epsilon_\ell(\nu)$ curves as those of the ``observed" star when considered as a function of frequency, but disagree as a function of radial order $n$. 
Conversely the ratios for the scaled model which has a different interior structure from that of the ``observed" star agree with the observations when considered as a function of $n$, but do not lie on the same 
$\epsilon_0(\nu) - \epsilon_\ell(\nu)$ curves as those of the ``observed" star a when considered as a function of frequency.

This demonstrates that seeking a best fit model by comparing separation ratios at the same $n$ values can give erroneous results.  

This is not surprising, the radial order $n$ of an eigenfrequency depends on the structure of the outer layers of a star as well as on the interior structure, being determined approximately by the integral number of wavelengths that fit into the acoustic radius of the star $T=\int dr/c$ where $c$ is the  sound speed. 
Since $c$ is small in the outer layers these layers can make a significant contribution to the acoustic radius. so the $n$ values are not independent of the structure of the outer layers; comparing observed and model separation ratios at the same $n$ does not subtract off the effect of the outer layers.

  \begin{figure}[t]
  \begin{center} 
   \includegraphics[width=8.9cm,height=9.9cm]{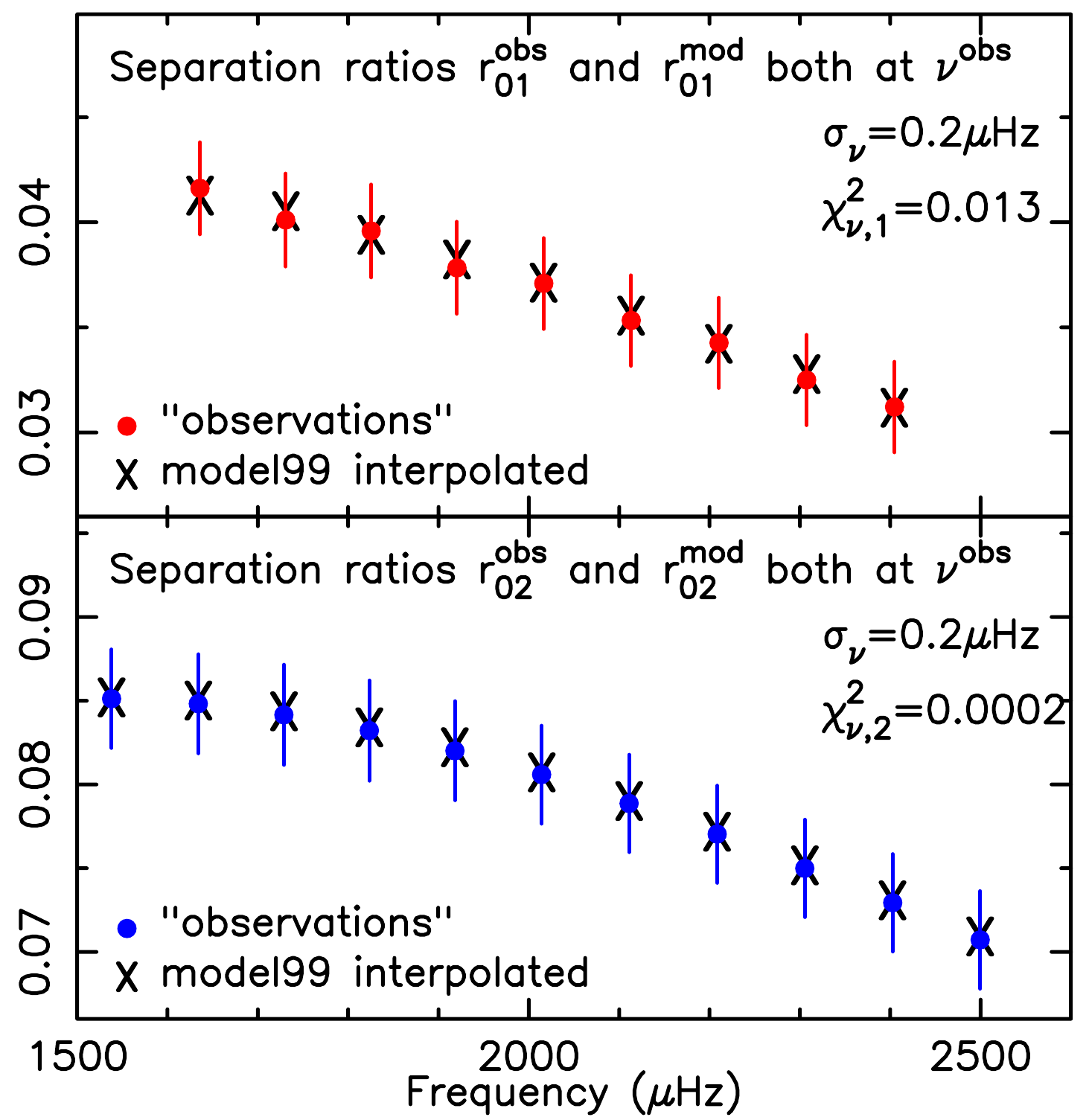}
  \end{center}  
   \vskip -15pt
   \caption{Comparison of ``observed" with model99 ratios interpolated to the ``observed" frequencies.}
\end{figure}

\section{Correcting the model fitting}
This comparison procedure can be corrected by interpolating the values of the model separation ratios $r_{0\ell}^{mod} (\nu^{mod}_{n\ell})$, which can be calculated for any $\{n,\ell\}$ values,  to derive values $r_{0\ell}^{mod} (\nu^{obs}_{n\ell})$ at the ``observed" frequencies, and then comparing the interpolated model ratios to the ``observed" ratios.  The results using local cubic interpolation on model99 ratios are shown in Fig. 5 along with the separate $\chi^2_{\nu}$ for the $r_{01}, r_{02}$  separations. The total $\chi^2_{\nu}$ 
$$\chi^2_{\nu} = {1\over N} \sum_\ell \sum_n  \left( r^{obs}_{0\ell}(\nu^{obs}_{n\ell}) - r^{mod}_{0\ell}(\nu^{obs}_{n\ell})\over \sigma^r_{0\ell}(\nu^{obs}_{n\ell})\right)^2
~~~~\ell=1,2\eqno(4)$$
is $0.006$ (with $\sigma_\nu=0.2\mu$Hz). The observed and interpolated model values are in excellent agreement.
This is the  procedure to be used in comparing observed and model separation ratios; it is independent of the $n$ values the astronomer assigns to the frequencies.

\section{Examples from model fitting}
The above example is perhaps a little extreme, although it illustrates the point we are making.  We therefore undertook a study which closely follows a realistic  model fitting procedure, taking our "observed" star and seeking  best fit models out of a large data base of models. The data base we use was kindly supplied by A Miglio (2012) and contains over $200,000$ models, computed using the CLES and OSC codes (Scuflaire et al 2008a,b).  The model set has masses in the range $1.04-1.5M_\odot$, $4$ values of the initial Hydrogen abundance $XH=0.684, 0.694, 0,704, 0.714$,  $4$ values of heavy element abundance $Z=0.015, 0.02, 0.025, 0.03$, $2$ values of the mixing length parameter $\alpha_{con}=1.705, 1.905$,  $2$ values of a chemical overshoot parameter $\alpha_{ov} =0, 0.2 H_p$ and  with and without diffusion. The models are evolved from the pre-main sequence to early shell burning.    Our model ``observed" star has similar physics with $
XH=0.7, Z=0.02, \alpha_{con}=1.6, \alpha_{ov}=0$ and no diffusion.

\begin{figure}[t]
  \begin{center} 
   \includegraphics[width=8.9cm]{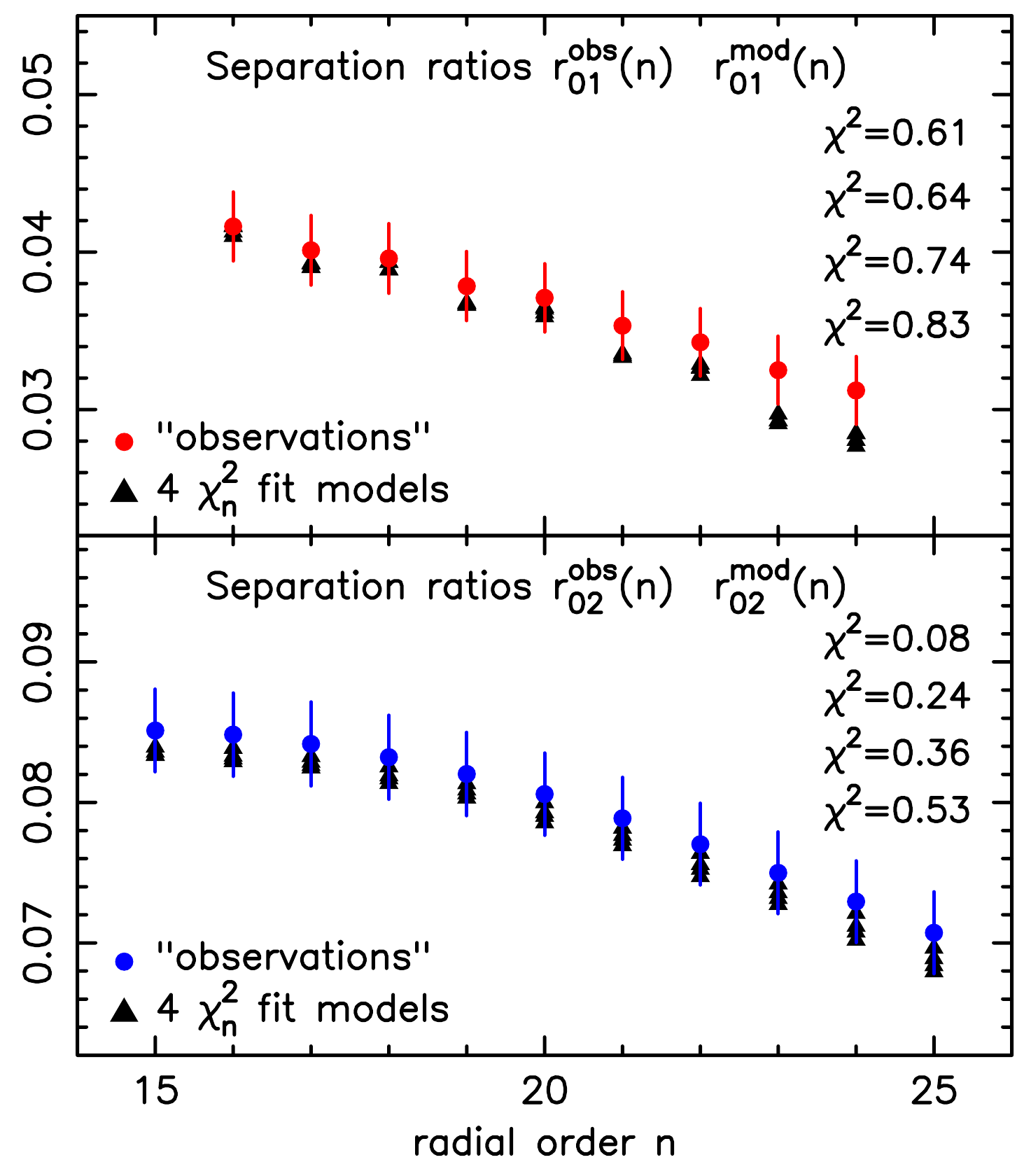}
  \end{center}  
   \vskip -15pt
   \caption{``Observed" and model ratios for the 4 $\chi^2_n$ models.}   
\end{figure}

We searched for best fit models that lie within an error box in the HR-diagram around the "observed" star  defined by  $T_{eff}=6242\pm 70^o$K, 
$L/L_\odot= 2.45\pm0.245$, these being realistic error estimates. 

Some $1,324$ models lie within the error box in the HR diagram of which $53$ have both  $\chi^2_{\nu}\le 1$ and $\chi^2_n\le 1$; $38$ have  $\chi^2_{\nu}\le 1$ and $\chi^2_n >  1$; and  $17$ have $\chi^2_{n}\le 1$ and $\chi^2_{\nu} > 1$.  The fact that a large proportion of models that have
 $\chi^2_\nu \le 1$  also have $\chi^2_n \le 1$ is not a great surprise;  if the $\epsilon_0(\nu)-\epsilon_\ell(\nu)$ are slowly varying functions of $\nu$, the  difference  
$$\chi^2_{n}-\chi^2_\nu \approx {1\over N} \sum_\ell \sum_n   \left( {1\over\sigma^r_{n\ell}}  {d(\epsilon_0-\epsilon_\ell)\over d\nu} \big[\nu^{mod}_{n\ell} -\nu^{obs}_{n\ell}\big] \right)^2~~~~~
\eqno(5)$$
so if $d(\epsilon_\ell-\epsilon_0)/d\nu$ is small, and the frequency offset $\nu^{mod}_{n\ell} -\nu^{obs}_{n\ell}$ is not too large, this difference is small.

To illustrate the point that matching observed and model ratios at the same $n$ values can lead to erroneous inferences we show in Fig. 6 the ratios for $4$ models that lie within the error box in the HR diagram and have a  combined $\chi^2_n$ of $0.32, 0.42, 0.53, 0.66$;  apparently good fits to the "observed" data (details of the models are given in Table 1, they all include diffusion, have the same heavy element abundance  $Z$ and similar ages).
However if we interpolate for the values of the ratios at the "observed" frequencies the fit is not so good - this is shown in Fig. 7; the  combined $\chi^2_\nu$ for the $4$ models  are $2.06, 2.45,  2.61, 2.10$ which for 20 ratios has a probability of being due to chance of less than $0.01- 0.35\%$..

\vskip 10pt
\centerline{Table 1}

\centerline{Properties of the 4 models with $\chi_n^2 <1, \chi^2_\nu >2$}
\vskip 3pt
\small
\tabskip=1 em plus 4 em minus 0.8 em
\halign to \hsize  {#\hfil&&\hfil$#$\cr
\parindent-0pt
\hskip-3pt\hbox{$M/M_\odot$}&\hbox{$XH~\,$}&\hbox{$Z~~~$}&\hbox{$\alpha_{con}\,$}&\hbox{$\alpha_{ov}$}&\hbox{$dif$}&\hbox{$Xc\,$}&\hbox{$age9$}&\hbox{$\Delta~~~$}&\hbox{$\chi^2_n~~~$}&\hbox{$\chi^2_\nu~~~$} \cr
\hbox{$1.28$}&\hbox{$0.704$}&\hbox{$0.025$}&\hbox{$1.705$}&\hbox{$0.0$}&\hbox{$yes$}&\hbox{$0.34$}&\hbox{$2.56$}&\hbox{$91.1$} &\hbox{0.32}&\hbox{2.06} \cr
\hbox{$1.23$}&\hbox{$0.684$}&\hbox{$0.025$}&\hbox{$1.705$}&\hbox{$0.0$}&\hbox{$yes$}&\hbox{$0.32$}&\hbox{$2.65$}&\hbox{$92.8$}&\hbox{0.66} &\hbox{2.10} \cr
\hbox{$1.26$}&\hbox{$0.694$}&\hbox{$0.025$}&\hbox{$1.705$}&\hbox{$0.0$}&\hbox{$yes$}&\hbox{$0.33$}&\hbox{$2.58$}&\hbox{$91.2$} &\hbox{0.42} &\hbox{2.45} \cr
\hbox{$1.24$}&\hbox{$0.684$}&\hbox{$0.025$}&\hbox{$1.705$}&\hbox{$0.0$}&\hbox{$yes$}&\hbox{$0.32$}&\hbox{$2.58$}&\hbox{$91.5$} &\hbox{0.53}&\hbox{2.61} \cr
}
\vskip10pt
\normalsize

\begin{figure}[t]
  \begin{center} 
  \vskip 3pt
   \includegraphics[width=8.7cm]{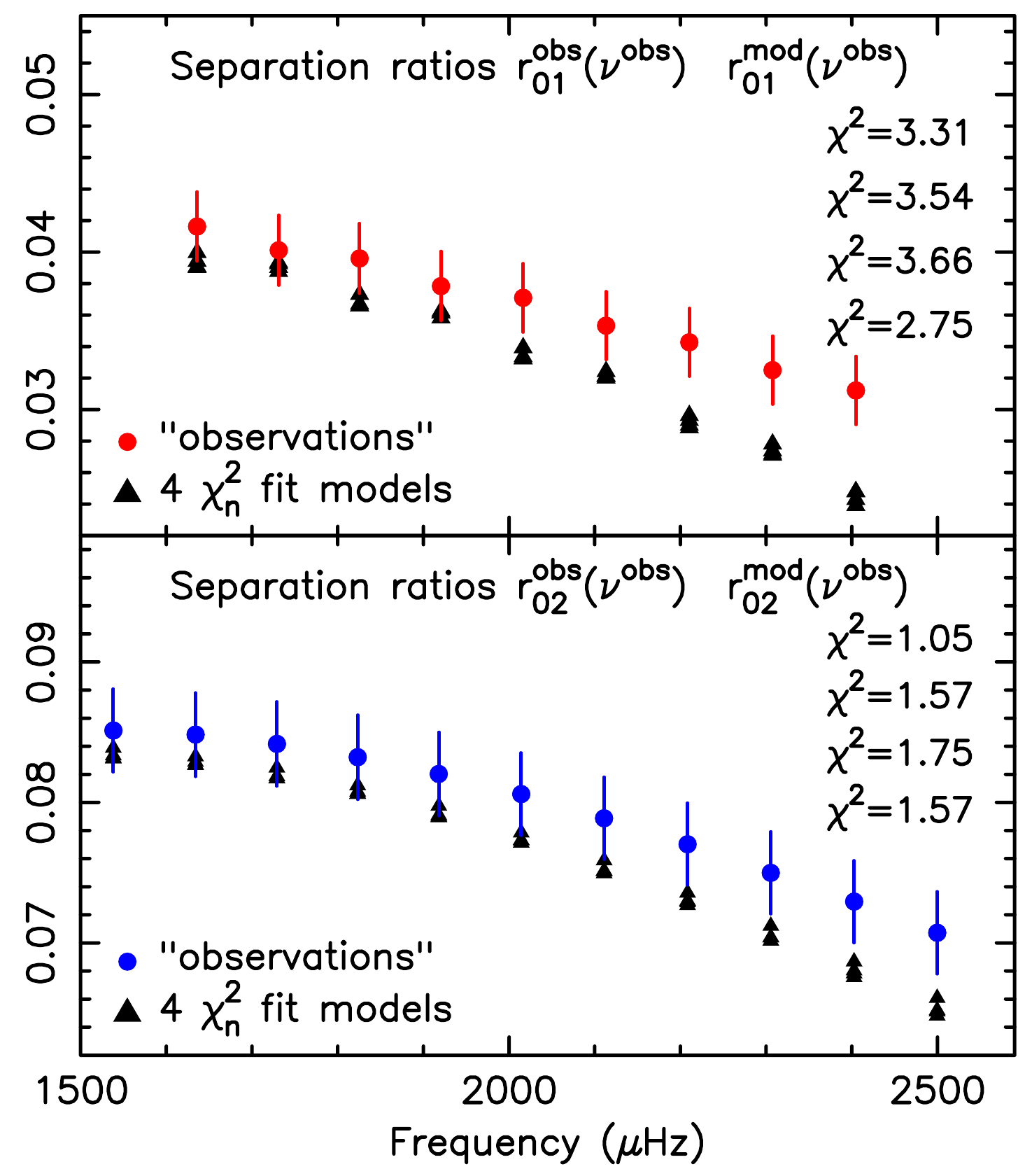}
  \end{center}  
   \vskip -15pt
   \caption{Separation ratios of the 4 $\chi^2_n$ models interpolated to the observed frequencies}   
   \vskip-10pt
\end{figure}
\begin{figure}[h]
  \begin{center} 
  \vskip 3pt
   \includegraphics[width=8.7cm]{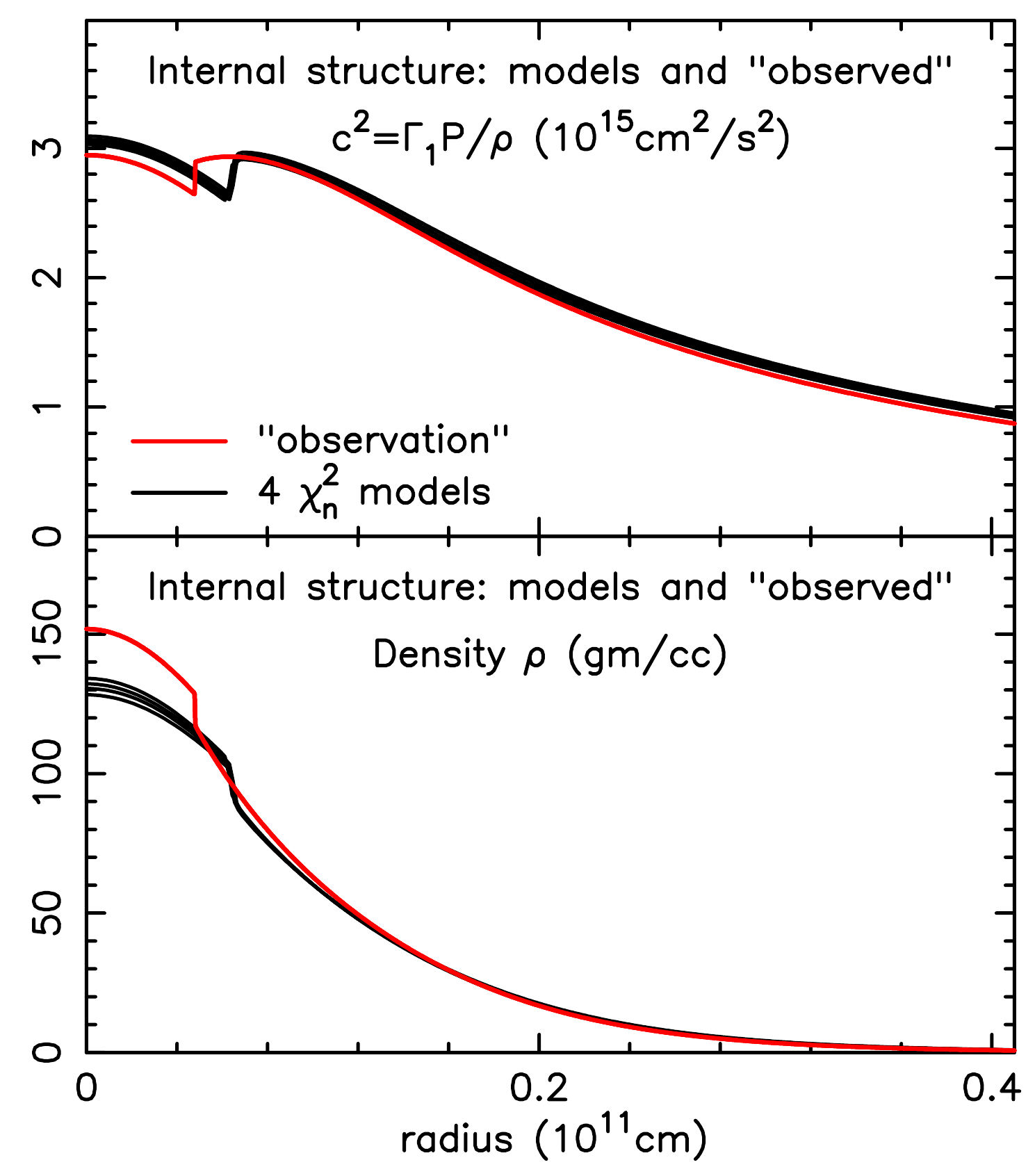}
  \end{center}  
   \vskip -15pt
   \caption{Internal structure the  ÒobservedÓ star and that of the 4 models obtained by fitting at the same $n$ values.}   
   \vskip-10pt
\end{figure}

In Fig. 8 we compare the sound speed and density of the 4 models in the inner layers  with that of the ``observed" star; the models have a higher central temperature, lower central density and larger convective core (the core boundary being at the point of increase in $c^2$). 
This demonstrates that models that fit the observed ratios when compared at the same $n$ values (within reduced $\chi^2 <\sim1$ for $\sigma_\nu=0.2\mu$Hz)  may not fit the observed ratios when compared at the observed frequencies and can have an interior structure that is not that of the observed star.

\begin{figure}[t]
  \begin{center} 
  \vskip 3pt
   \includegraphics[width=8.7cm]{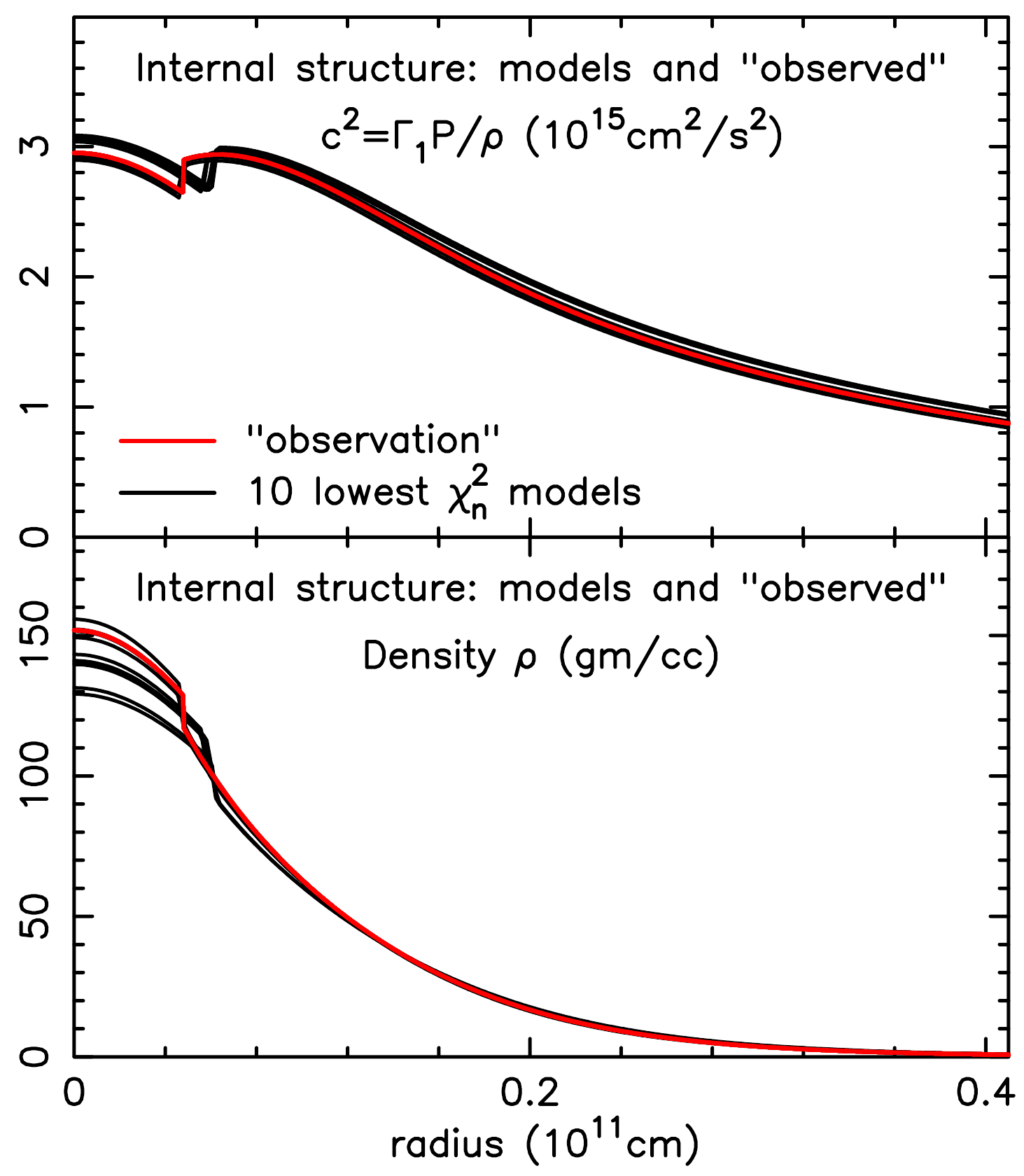}
  \end{center}  
   \vskip -15pt
   \caption{Internal structure the  ÒobservedÓ star and that of the best 10 models obtained by fitting at the same $n$ values.}   
   \vskip-10pt
\end{figure}

The above examples were chosen to illustrate the argument advanced in this paper - but they are not the models with the lowest reduced $\chi_n^2$.  The $10$ best fit models when compared at the same $n$ values have a combined reduced  $\chi^2_n  < 0.18$;  details of these models are given in Table 2 and their inner structure is shown in Fig. 9.  $4$ of the $10$ models fit the structure very well; but all $4$ also have a very small reduced $\chi^2_\nu$, and are amongst the 10 best fit models obtained by  comparing at the observed frequencies (see Table 3 below).  As pointed out above the difference between $\nu$ and $n$ comparison (Eqn. 4) can be small.  The other $6$ models,  which includes the model with the lowest $\chi^2_n$,  are not such a good fit,  having a larger convective core, lower central density and higher central temperature. 

Were one to seek a best fit model by comparing ratios at the same $n$ values, the model with the lowest $\chi^2_n$ would not necessarily  reproduce the internal structure of the observed star.   Matching ratios at the same $n$ values can therefore  give erroneous results on the  interior structure and properties of an observed star, and on the physical processes that govern stellar evolution.

\centerline{Table 2}

\centerline{Properties of the 10 models with the lowest $\chi_n^2$}
\small
\vskip 3pt
\tabskip=1 em plus 4 em minus 0.8 em
\halign to \hsize  {#\hfil&&\hfil$#$\cr
\parindent-0pt
\hskip-3pt\hbox{$M/M_\odot$}&\hbox{$XH~\,$}&\hbox{$Z~~~$}&\hbox{$\alpha_{con}\,$}&\hbox{$\alpha_{ov}$}&\hbox{$dif$}&\hbox{$Xc\,$}&\hbox{$age9$}&\hbox{$\Delta~~~$}&\hbox{$\chi^2_n~~~$} \cr
\hbox{$1.19$}&\hbox{$0.684$}&\hbox{$0.020$}&\hbox{$1.705$}&\hbox{$0.0$}&\hbox{$yes$}&\hbox{$0.32$}&\hbox{$2.50$}&\hbox{$94.8$} &\hbox{0.089} \cr
\hbox{$1.18$}&\hbox{$0.694$}&\hbox{$0.020$}&\hbox{$1.705$}&\hbox{$0.0$}&\hbox{$no\,$}&\hbox{$0.30$}&\hbox{$2.81$}&\hbox{$99.8$} &\hbox{0.107} \cr
\hbox{$1.22$}&\hbox{$0.694$}&\hbox{$0.020$}&\hbox{$1.705$}&\hbox{$0.0$}&\hbox{$yes$}&\hbox{$0.33$}&\hbox{$2.46$}&\hbox{$92.6$} &\hbox{0.123} \cr
\hbox{$1.22$}&\hbox{$0.714$}&\hbox{$0.020$}&\hbox{$1.705$}&\hbox{$0.0$}&\hbox{$no\,$}&\hbox{$0.30$}&\hbox{$2.90$}&\hbox{$97.6$}&\hbox{0.144} \cr
\hbox{$1.20$}&\hbox{$0.684$}&\hbox{$0.020$}&\hbox{$1.705$}&\hbox{$0.0$}&\hbox{$yes$}&\hbox{$0.33$}&\hbox{$2.43$}&\hbox{$93.4$} &\hbox{0.151} \cr
\hbox{$1.28$}&\hbox{$0.704$}&\hbox{$0.025$}&\hbox{$1.705$}&\hbox{$0.0$}&\hbox{$no\,$}&\hbox{$0.34$}&\hbox{$2.55$}&\hbox{$93.2$}&\hbox{0.156} \cr
\hbox{$1.20$}&\hbox{$0.704$}&\hbox{$0.020$}&\hbox{$1.705$}&\hbox{$0.0$}&\hbox{$no\,$}&\hbox{$0.30$}&\hbox{$2.89$}&\hbox{$98.4$}&\hbox{$0.156$}\cr
\hbox{$1.21$}&\hbox{$0.694$}&\hbox{$0.020$}&\hbox{$1.705$}&\hbox{$0.0$}&\hbox{$yes$}&\hbox{$0.33$}&\hbox{$2.48$}&\hbox{$94.5$}&\hbox{0.159} \cr
\hbox{$1.26$}&\hbox{$0.694$}&\hbox{$0.025$}&\hbox{$1.705$}&\hbox{$0.0$}&\hbox{$no\,$}&\hbox{$0.33$}&\hbox{$2.57$}&\hbox{$93.4$} &\hbox{0.175} \cr
\hbox{$1.16$}&\hbox{$0.684$}&\hbox{$0.020$}&\hbox{$1.705$}&\hbox{$0.0$}&\hbox{$no\,$}&\hbox{$0.28$}&\hbox{$2.86$}&\hbox{$100.1$} &\hbox{0.179} \cr
}
\vskip10pt
\normalsize

\begin{figure}[t]
  \begin{center} 
   \includegraphics[width=8.9cm]{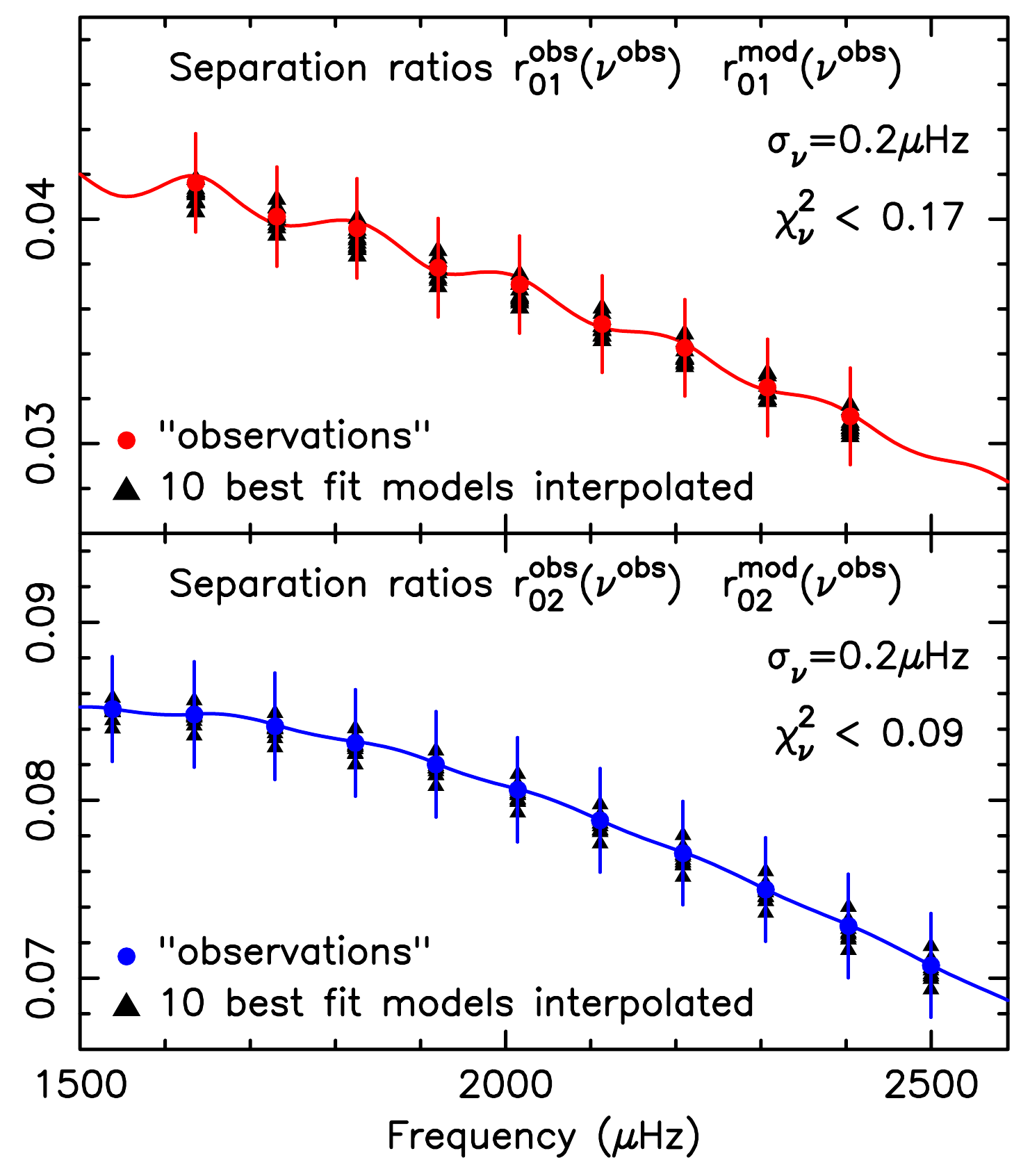}
  \end{center}  
   \vskip -15pt
   \caption{Comparison of ÒobservedÓ ratios with those of the best 10 models interpolated to the ÒobservedÓ frequencies.}   
   \vskip-3pt
\end{figure}
\begin{figure}[h]
  \begin{center} 
   \includegraphics[width=8.9cm]{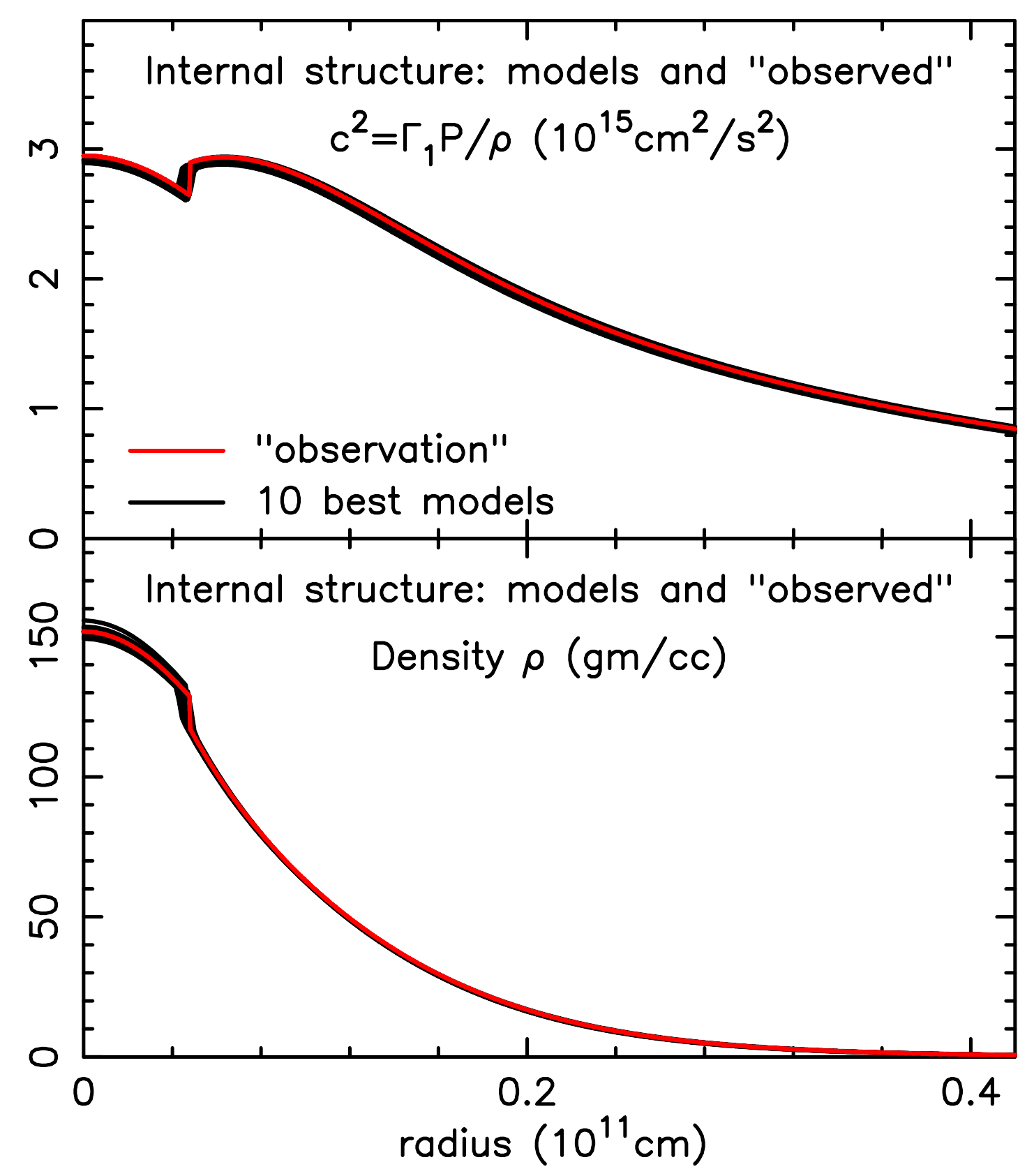}
  \end{center}  
   \vskip -15pt
   \caption{Comparison of the internal structure ÒobservedÓ star with those of the best 10 models.}   
   \vskip-10pt
\end{figure}
The $10 $ best fit models when comparing ratios interpolated to the observed frequencies are illustrated in Fig. 10; the fit is very good with
$\chi^2_\nu< 0.132$ for all the models.    The internal structure of these models is shown in Fig. 11;  all the models satisfactorily reproduce the internal structure of the  ``observed" star. 

Table 3 lists the properties of these $10$  models; they have a range of masses, initial hydrogen abundance and some include diffusion and some do not. 
The best fit model having $\chi^2_\nu=0.013$ is as close as it could be to the ``observed" star given that the model set does not contain models with exactly the same parameters $XH,  \alpha_{con}$  as the "observed" star.  Models $1,  4, 5, 7$ in Table 3 are also present in Table 2 and are the 4 models that fit the internal structure of the observed star in Fig. 8.
\vskip 10pt
\centerline{Table 3}

\centerline{Properties of the 10 models with the lowest $\chi_\nu^2$}
\vskip 3pt
\small
\tabskip=1 em plus 4 em minus 0.8 em
\halign to \hsize  {#\hfil&&\hfil$#$\cr
\hskip-3pt\hbox{$M/M_\odot$}&\hbox{$XH~\,$}&\hbox{$Z~~~$}&\hbox{$\alpha_{con}\,$}&\hbox{$\alpha_{ov}$}&\hbox{$dif$}&\hbox{$Xc\,$}&\hbox{$age9$}&\hbox{$\Delta~~~$}&\hbox{$\chi^2_\nu~~~$} \cr
\hbox{$1.20$}&\hbox{$0.704$}&\hbox{$0.020$}&\hbox{$1.705$}&\hbox{$0.0$}&\hbox{$no\,$}&\hbox{$0.30$}&\hbox{$2.89$}&\hbox{$98.4$} &\hbox{0.013} \cr
\hbox{$1.17$}&\hbox{$0.694$}&\hbox{$0.020$}&\hbox{$1.905$}&\hbox{$0.0$}&\hbox{$yes$}&\hbox{$0.30$}&\hbox{$2.87$}&\hbox{$102.9$} &\hbox{0.018} \cr
\hbox{$1.18$}&\hbox{$0.704$}&\hbox{$0.020$}&\hbox{$1.905$}&\hbox{$0.0$}&\hbox{$yes$}&\hbox{$0.30$}&\hbox{$2.95$}&\hbox{$103.6$} &\hbox{0.030} \cr
\hbox{$1.22$}&\hbox{$0.714$}&\hbox{$0.020$}&\hbox{$1.705$}&\hbox{$0.0$}&\hbox{$no\,$}&\hbox{$0.30$}&\hbox{$2.90$}&\hbox{$97.6$} &\hbox{0.032} \cr
\hbox{$1.16$}&\hbox{$0.684$}&\hbox{$0.020$}&\hbox{$1.705$}&\hbox{$0.0$}&\hbox{$no\,$}&\hbox{$0.28$}&\hbox{$2.86$}&\hbox{$100.1$}&\hbox{0.039} \cr
\hbox{$1.18$}&\hbox{$0.704$}&\hbox{$0.020$}&\hbox{$1.705$}&\hbox{$0.0$}&\hbox{$yes$}&\hbox{$0.30$}&\hbox{$2.96$}&\hbox{$99.7$} &\hbox{0.058} \cr
\hbox{$1.18$}&\hbox{$0.694$}&\hbox{$0.020$}&\hbox{$1.705$}&\hbox{$0.0$}&\hbox{$no\,$}&\hbox{$0.30$}&\hbox{$2.81$}&\hbox{$99.8$}&\hbox{$0.063$} \cr
\hbox{$1.20$}&\hbox{$0.714$}&\hbox{$0.020$}&\hbox{$1.905$}&\hbox{$0.0$}&\hbox{$yes$}&\hbox{$0.30$}&\hbox{$2.99$}&\hbox{$102.6$}&\hbox{0.074} \cr
\hbox{$1.20$}&\hbox{$0.714$}&\hbox{$0.020$}&\hbox{$1.705$}&\hbox{$0.0$}&\hbox{$yes$}&\hbox{$0.31$}&\hbox{$2.92$}&\hbox{$99.5$} &\hbox{0.080} \cr
\hbox{$1.17$}&\hbox{$0.694$}&\hbox{$0.020$}&\hbox{$1.705$}&\hbox{$0.0$}&\hbox{$yes$}&\hbox{$0.29$}&\hbox{$2.92$}&\hbox{$98.5$} &\hbox{0.132} \cr
}
\vskip10pt
\normalsize
\vskip-10pt


\begin{figure}[h]
  \begin{center} 
  \vskip 3pt
   \includegraphics[width=8.7cm,height=6cm]{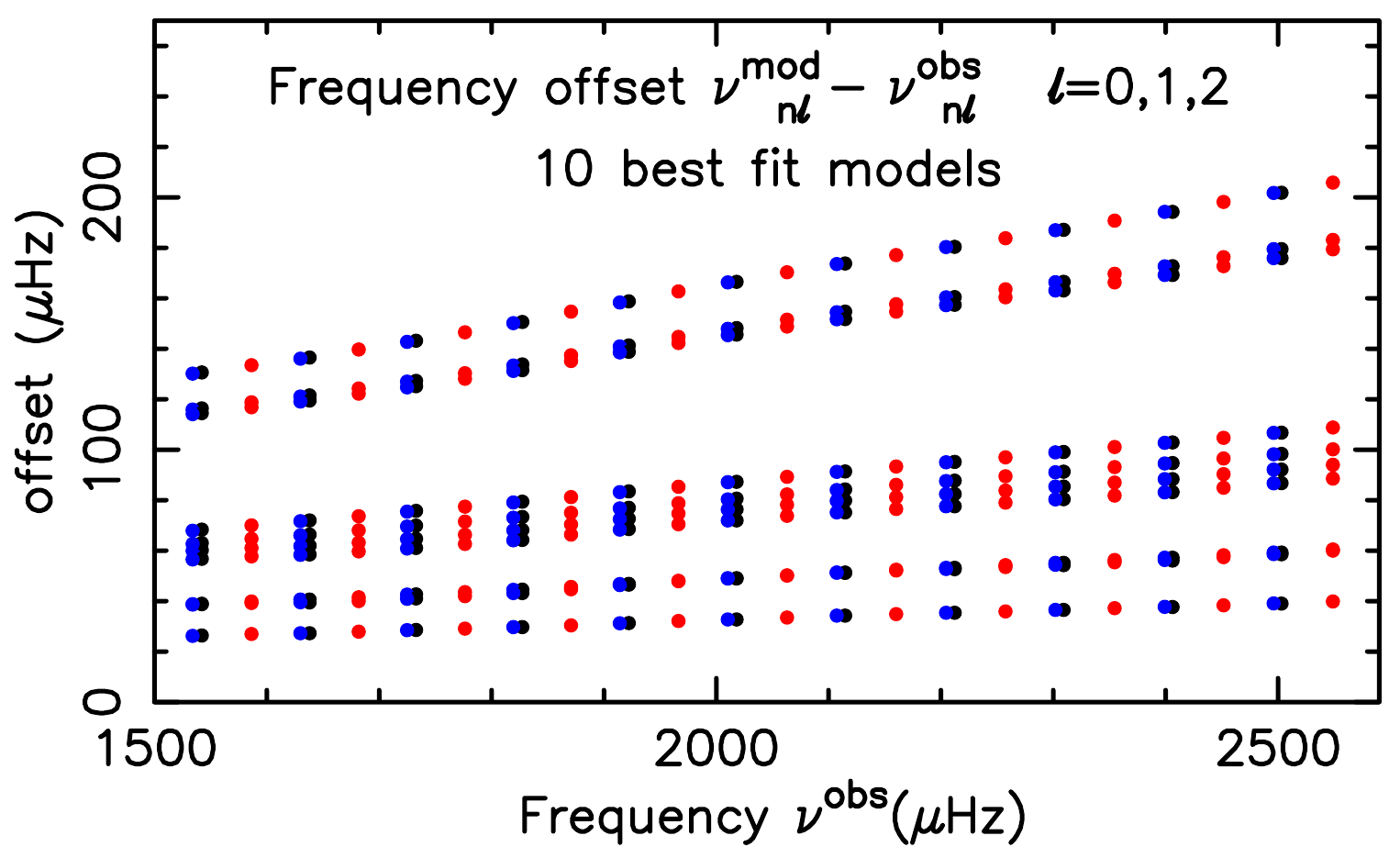}
  \end{center}  
   \vskip -15pt
   \caption{Frequency offset of frequencies of the $10$ best fit models from those of the ``observed" star.}   
\end{figure}
\eject
The difference between the observed frequencies $\nu^{obs}$ and the model frequencies $\nu^{mod}_{n\ell}$, often referred to as the ``frequency offset" is plotted in Fig. 12. These offsets can be extremely large even though the interior structure of the models agrees well with that of the ``observed star.


\section{Separation ratios  and internal phase shifts}
The frequencies of a spherical star satisfy  an Eigenfrequency equation of the form (Roxburgh and Vorontsov 2000)
$$2\pi T \nu_{n\ell}= \pi\, \big[n+\ell/2 + \alpha_\ell(\nu) -\delta_\ell(\nu)\big],~~~{\rm where}~~T=\int_0^R{dr\over c}\eqno(6)$$
is the acoustic radius of the star and $c$ the sound speed.  The {\it inner phase shift} $\delta_\ell(\nu)$ is determined by the structure interior to some fitting point $r_f$,  and  {\it outer phase shifts} $\alpha_\ell(\nu)$ by  the structure above $r_f$.  
This is exactly true for modes of degree $\ell=0,1$ where the oscillation equations reduce to $2^{nd}$ order (Takata 2005, Roxburgh 2006,8b), and holds to high accuracy for $\ell=2$ provided $r_f$ is in the low density outer layers of the star. 

Equation 6  is identical in form to that of Eqn. 3 with $\epsilon_\ell=\alpha_\ell-\delta_\ell$ and ${\bar\Delta}=1/(2T)$. 
The important difference is that since the density, sound speed and curvature effects are small in the outer layers, the $\alpha_\ell(\nu)$ are almost independent of degree $\ell$.

Following Roxburgh and Vorontsov (2000) we define $\alpha_\ell(\nu, t), \delta_\ell(\nu,t)$  at any radius and frequency in terms of the scaled Eulerian pressure perturbation $\psi_\ell(\nu, t) =r p'/(\rho c)^{1/2}$ by
$$\tan[2\pi\nu\,t -\ell\pi/2+ \pi \delta_\ell(\nu, t)] = {2\pi\nu\psi\over d\psi/dt} ~~~~~~~~t =\int_0^r {dr\over c(r)} $$
$$\tan[2\pi\nu\,\tau - \pi \alpha_\ell(\nu, \tau)] = {2\pi\nu\psi\over d\psi/d\tau} ~~~~~~~~~~~~~~\tau=\int_R^r {dr\over c(r)} \eqno(7)$$
where $t$ is the acoustic radius, $\tau=T-t$ the acoustic depth, and $\nu$ any frequency (not restricted to an eigenfrequency). The  
eigenfunctions $\psi$ are then almost pure sine waves with constant phase shift in the outer layers and the phase shifts are almost independent of the choice of fitting point $r_f$ (or $t_f$) (Roxburgh \& Vorontsov 1996). 

In Fig. 13 we show the differences $\alpha_2(\nu)-\alpha_0(\nu)$ for the ``observed" $1.2M_\odot$ star, evaluated at  radii {$x_f=r_f/R=0.93, 0.95,$ $ 0.97$. [ Note $\alpha_1-\alpha_0 \approx (\alpha_2-\alpha_0)/3$.]  Since $2.28<\alpha_0(\nu) < 3.15$ for $1500<\nu<2500\mu$Hz the dependence of $\alpha_\ell$ on $\ell$ is negligible.

\begin{figure}[h]
  \begin{center} 
   \includegraphics[width=8.9cm]{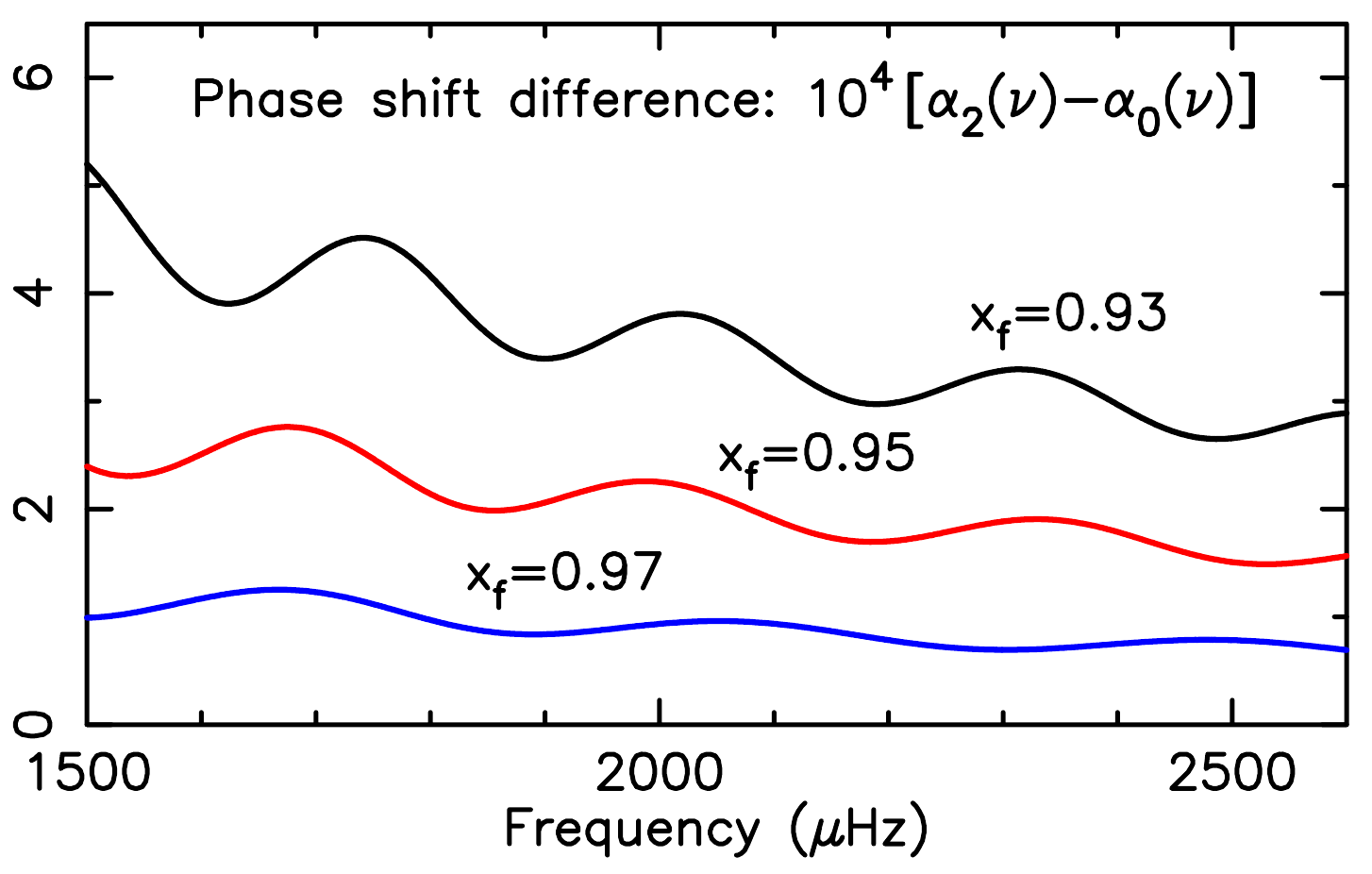}
  \end{center}  
   \vskip -6pt
   \caption{Phase shift differences $\alpha_2(\nu)-\alpha_0(\nu)$ at fractional radii $x_f=0.93, 0.95,  0.97$ for the model ``observed" star'. }
\end{figure}

\eject
 With $\alpha_\ell(\nu)$ independent of $\ell$ it follows that
$$\delta_\ell(\nu)-\delta_0(\nu) =\epsilon_0(\nu)-\epsilon_\ell(\nu) \eqno(8)$$
which is determined by the interior structure of the star. 
In Fig. 14 we show the differences $\delta_\ell(\nu)-\delta_0(\nu)$ for the ``observed" $1.2M_\odot$ star, evaluated at  radii $x_f=r_f/R=0.93,0.95,0.97$. One can hardly  distinguish the curves for $\delta_\ell-\delta_0$ which demonstrates their  insensitivity to the location of the fitting point.   

\begin{figure}[t]
  \begin{center} 
   \includegraphics[width=8.9cm]{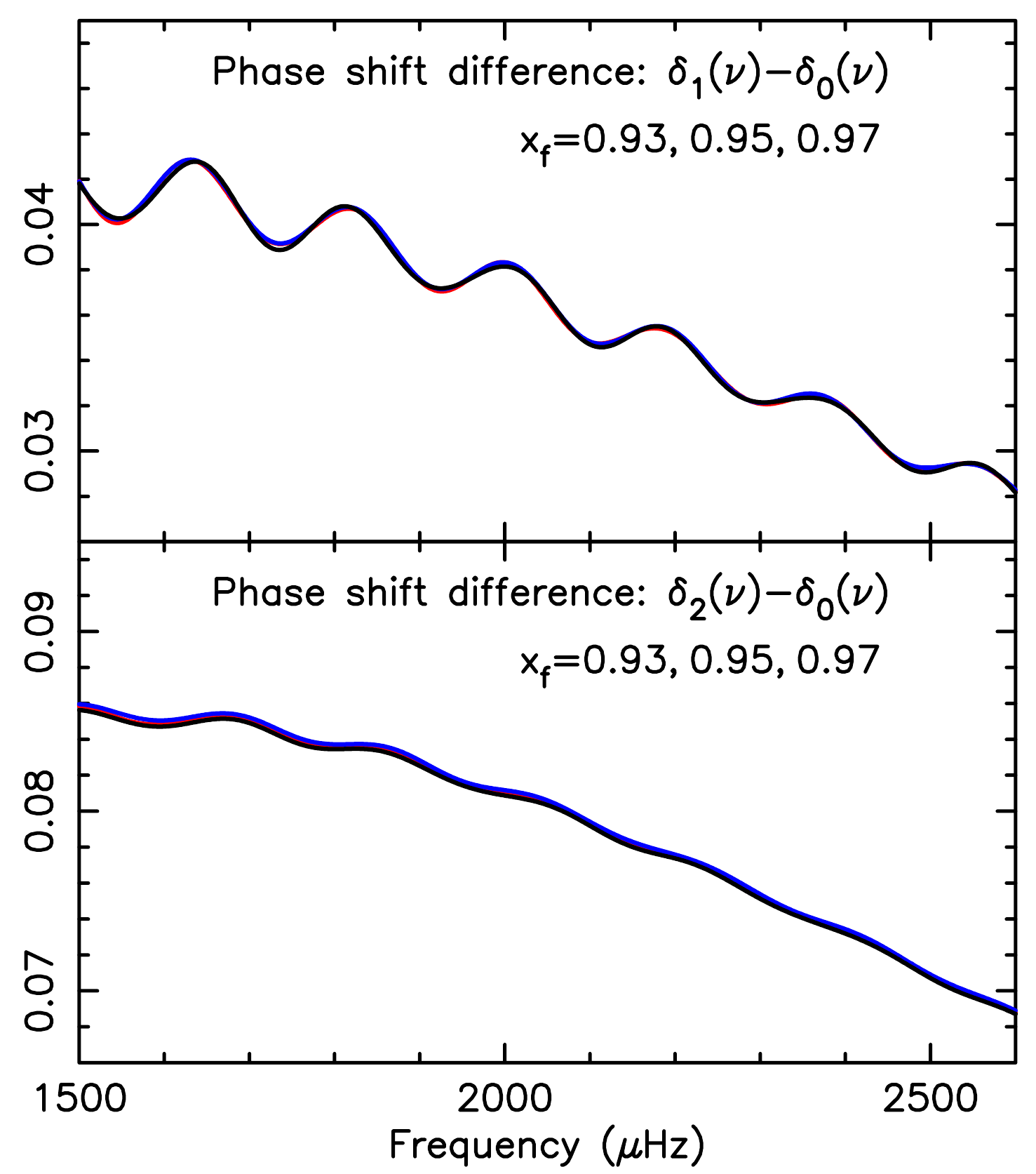}
  \end{center}  
   \vskip -6pt
   \caption{ Phase shift differences $\delta_\ell(\nu)-\delta_0(\nu)$ at fractional radii $x_f=0.93, 0.95,  0.97$ for the model ``observed" star'; top panel: 
$\delta_1(\nu)-\delta_0(\nu)$; bottom panel: $\delta_2(\nu)-\delta_0(\nu)$.   
   }
\end{figure}

\section{Conclusions and  comments on model fitting}
The separation ratios are an approximation to the phase shift differences $\epsilon_0(\nu)-\epsilon_\ell(\nu)$ which are determined by the  structure of the stellar interior and are almost independent of the structure of the surface layers. One should not compare  the ratios $r^{obs}_{0\ell}(n)$ of an observed frequency set with the $r^{mod}_{0\ell}(n)$ of a model at the same $n$ values since, unless the frequencies coincide (in which case no comparison is needed !)  one is not comparing values of $\epsilon_0-\epsilon_\ell$ at the same frequencies. This can be corrected by interpolating in the  model values $r^{mod}_{0\ell}(n)$ to determine the ratios at the observed frequencies $r^{mod}_{0\ell}(\nu^{obs})$ and then comparing observed and model values at the observed frequencies.

The error in comparing ratios at the same $n$ values depends both on the scale of variation of $\epsilon_0-\epsilon_\ell$ and the difference in frequencies at the same $n$ values (Eqn. 4).  For the model fitting examples given here, with an error on the frequencies of $0.2\mu$Hz, the difference is large enough to lead to erroneous conclusions on the structure, age, and physical processes that govern stellar evolution. 

Note that instead of using separation ratios one could interpolate the phase shifts $\epsilon_\ell$ at (for example) $\nu^{obs}_{n0}$ for both the observations and models and directly compare their respective phase shift differences   $\epsilon_0-\epsilon_\ell$ at $\nu^{obs}_{n0}$.

A further comment is in order. Since the objective of using separation ratios is to subtract off the unknown effect of the outer layers of a star it can only yield a best fit to the interior structure.  it is inconsistent to impose strict constraints on the radius, effective temperature, surface gravity, metallicity, and large separations derived by observations, since these all depend on the structure of the outer layers and the effects of diffusion.   The luminosity on the other hand is determined by the interior structure alone, as is the mass which, in principle, can be estimated from surface gravity  and surface radius, and these can provide additional constraints on the model fitting. 

Finally we add a note of caution. The comparison of separation ratios, or phase shift differences, depends on interpolation, and interpolation is not an exact science. However if one uses the same procedure for both observations and models then if one finds a good fit this should be almost independent of the interpolation procedure used. 

\section*{Acknowledgements}
We thank Andrea Miglio who generously provided the large data base of models used in the model fitting.   I W Roxburgh gratefully acknowledges support from the Leverhulme Foundation under grant EM-2012-035/4

\newpage
 \end{document}